\documentclass[useAMS,usenatbib]{mn2e}

\usepackage{graphicx}
\usepackage{natbib}

\newcommand\aap{A\&A}

\newcommand\aj{AJ}
\newcommand\mnras{MNRAS}
\newcommand\apj{ApJ}

\newcommand\ibvs{IBVS}

\title[V821 Cas]{Absolute parameters of the eclipsing binary V821 Cas from UBVRI light curves and radial velocities.\thanks{Based on 
  observations collected at Catania Astrophysical Observatory (Italy) and T\"{U}B\.ITAK National 
  Observatory (Antalya, Turkey).} \thanks{{Table 1 is only available in electronic form at the CDS via anonymous ftp to http://www.blackwell-syngery.com/doi....}}
  }
\author[\"{O}. \c{C}ak{\i}rl{\i} et al.]
{\"{O}. \c{C}ak{\i}rl{\i}$^1$\thanks{E-mail: omur.cakirli@ege.edu.tr }, C.~\.{I}bano\v{g}lu$^1$, S.~Bilir$^2$, E. Sipahi$^1$ \\
$^1$Ege University, Science Faculty, Department of Astronomy and Space Sciences, 35100 Bornova, \.{I}zmir, Turkey\\
$^2$\.{I}stanbul University, Science Faculty, Department of Astronomy and Space Sciences, 34119 University, \.{I}stanbul, Turkey\\
}
\begin{document}

\date{Accepted 2008 Month Day. Received 2008 Month Day; in original form 2008 ??? ?}

\pagerange{\pageref{firstpage}--\pageref{lastpage}} \pubyear{2007}

\maketitle

\label{firstpage}

\begin{abstract}
We present UBVRI photometric measurements and spectroscopic observations of the double-lined eclipsing binary V821\,Cas. The
radial velocities were obtained by means of the cross-correlation technique. Simultaneous analyses of the multi-band light curves and RVs give
the absolute parameters for the stars as: 
M$_1$=2.05$\pm$0.07  M$_{\odot}$, M$_2$=1.63$\pm$0.06 M$_{\odot}$, R$_1$=2.31$\pm$0.03 R$_{\odot}$, 
R$_2$=1.39$\pm$0.02 R$_{\odot}$, T$_{eff_1}$=9\,400$\pm$400 K, and T$_{eff_2}$=8\,600$\pm$400 K. Analysis of the O-C residuals 
yielded an apsidal motion in the binary at a rate of $\dot{\omega}$=0$^{\degr}$.0149$\pm$0$^{\degr}$.0023 cycle$^{-1}$, corresponding to 
an apsidal period of U=118$\pm$19 yr. Subtracting the relativistic contribution we find that $\log~k_{2obs}$=-2.590 which is in agreement with 
the value predicted by theoretical models. Comparison with current stellar evolution models gives an age of $5.6\times10^{8}$ yr for the system.
\end{abstract}


\section{Introduction}
\label{}
V821\,Cas (BD +52$^{\circ}$ 3571, Tycho 4001-1445-1, V=8$^m$.31, (B-V)=0.11) was discovered to be an eclipsing binary of early A spectral
type, with a variability period of $\sim$1.8 days by the Hipparcos satellite (ESA, 1997). De\v{g}irmenci et al. 
(2003, 2007) determined an improved light curve and listed 25 photographic times of minima from which they derived an improved
eclipse ephemeris. They obtained {\em BVR} light curves, and established that the orbit is eccentric from the displacement of 
the secondary minimum. Its components do not show intrinsic variability, they are well-separated, and well within their Roche
lobes, which values them proper tests for stellar models. 

Several times of minima for V821\,Cas have appeared in the literature since, but no radial velocity measurements and multi-band 
light curves have been published. The aim of this work is to obtain the absolute parameters of the system. Along the next 
section we first present our spectroscopic data and the set of {\em UBVRI} light curves. Making use of the spectra we determine 
the parameters of the radial velocity curve and perform a reliable estimation of the spectral types for both 
components. Then, using the results from the spectroscopy together with the our photometric data we analyze the light 
curves to obtain reliable solutions that allow to determine the rest of the parameters of the system.

\section{Observations}
\subsection{Photometric observations}
We report here new photometry of V821\,Cas in the Bessell {\em UBVRI} bands. The photometric accuracy and the phase coverage 
(over 350 observations) are sufficient to guarantee a reliable determination of the light curve parameters thus
permitting a critical evaluation of stellar models. The observations were carried out with the 0.40-m telescope located on 
Mt. Bak{\i}rl{\i}tepe in September of 2007 at the T\"{U}B\.{I}TAK National Observatory (TUG, located in south of Turkey). The 
telescope is equipped with a Apogee 1k$\times$1k CCD (binned 2$\times$2) and standard Bessel {\em UBVRI} filters. 

The instrument with attached camera provide a field-of-view of 11$^{\prime}$.3$\times$11$^{\prime}$.3. 
By placing V821\,Cas very close to center of the CCD to get highest accuracy, we managed to strategically locate 
the binary on the chip together with two other stars of similar apparent magnitude. As comparison star we selected 
BD +52$^{\circ}$\,3575. The check star chosen to be BD +52$^{\circ}$\,3580. Both stars passed respective tests for 
intrinsic photometric variability and proved to be stable during time span of our observations. 

We collected a total of 1600 points in the UBVRI bands. The resulting V-band magnitude differences (var-comp) are listed
in Table 1 (available in electronic form at the CDS). A typical precision of the differential magnitudes 
is about 0.008 mag per measurement. Standard 
IRAF\footnote{IRAF is distributed by the National Optical Observatory, which is operated by the Association of 
the Universities for Research in Astronomy, inc. (AURA) under cooperative agreement with the National Science 
Foundation} tasks were used to remove the electronic bias and to perform the flat-fielding corrections. The IRAF 
task {\sc imalign} was used to remove the differences in the pixel locations of the stellar images and to place 
all the CCD images on the same relative coordinate systems. The data were analyzed using another IRAF task {\sc phot}
with no differential extinction effects taken into account given the relative small separation between the target and the comparison
and check stars. The phase-folded light curves for the whole observations are shown in Fig. 5 with all bands.

\begin{figure}
\includegraphics[width=10cm]{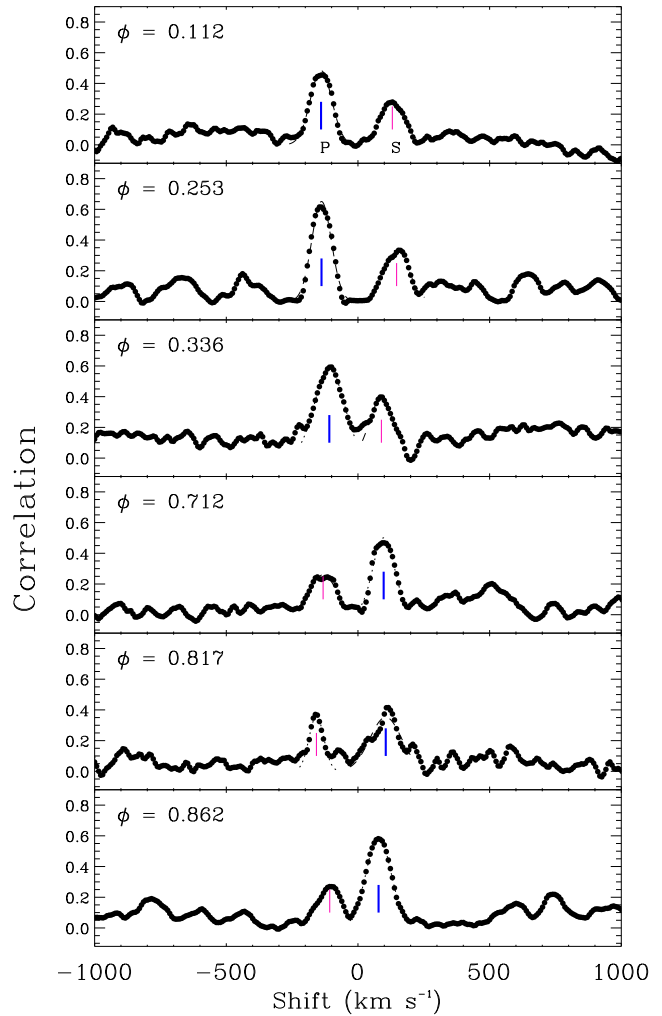}
\caption{Sample of Cross Correlation Functions (CCFs) between V821\,Cas and the radial velocity template 
spectrum (Vega) around the first and second quadrature.}
\label{ccf;fig1}
\end{figure}

\subsection{Spectroscopic observations}
Spectroscopic observations have been performed with the \'{e}chelle spectrograph (FRESCO) at the 91-cm
telescope of Catania Astrophysical Observatory. The spectrograph is fed by the telescope through an optical 
fibre ($UV$--$NIR$, 100 $\mu$m core diameter) and is located, in a stable position, in the room below the dome 
level. Spectra were recorded on a CCD camera equipped with a thinned back--illuminated SITe CCD of 
1k$\times$1k pixels (size 24$\times$24 $\mu$m). The cross-dispersed \'{e}chelle configuration yields a resolution 
of about 20\,000, as deduced from the full width at half maximum of the lines of the Th--Ar calibration lamp. The 
spectra cover the wavelength range from 4300 to 6650 {\AA}, split into 19 orders. In this spectral region, and 
in particular in the blue portion of the spectrum, there are several lines useful for the measure of radial velocity, as 
well as for spectral classification of the stars.

The data reduction was performed by using the \'{e}chelle task of IRAF package following the standard steps: background 
subtraction, division by a flat field spectrum given by a halogen lamp, wavelength calibration using the emission lines of
a Th-Ar lamp, and normalization to the continuum through a polynomial fit. 

Sixteen spectra of V821\,Cas were collected during the 20 observing nights between August 2 and 23, 2006. Typical
exposure times for the V821\,Cas spectroscopic observations were between 2400 and 2600 s. The signal-to-noise
ratio ($S/N$) achieved was between 70 and 115, depending on atmospheric condition. $\alpha$ Lyr (A0V), 59 Her
(A3IV), $\iota$ Psc (F7V), HD 27962 (A2IV), and $\tau$ Her were observed during each run as radial velocity and/or 
rotational velocity templates. The average $S/N$ at continuum in the spectral region of interest was 150--200 for the 
standard stars.

\begin{figure*}
\includegraphics[width=14cm]{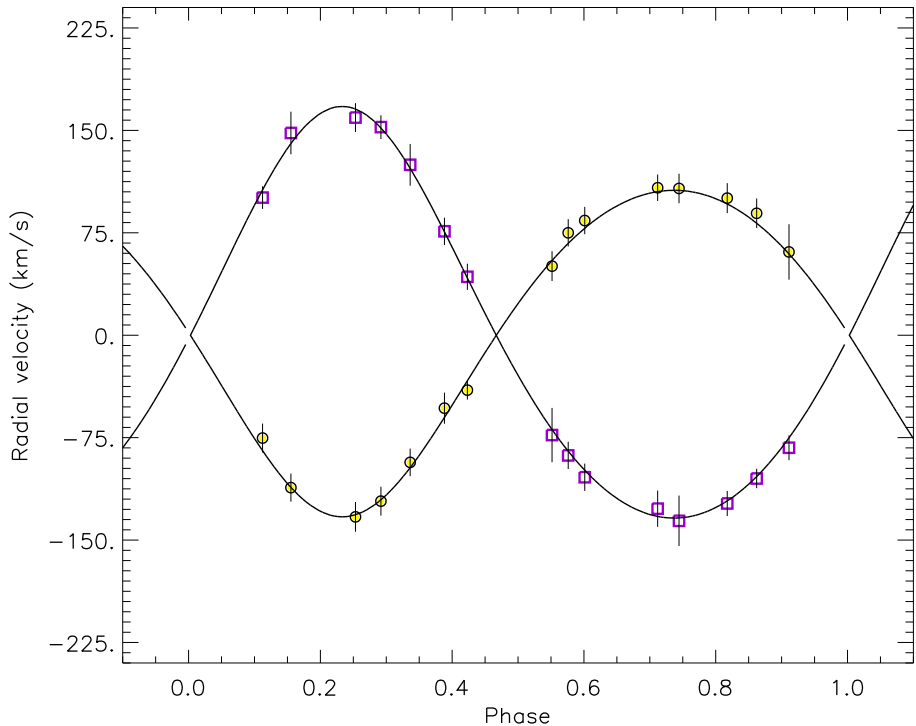}
\caption{Radial velocity curve folded on a period of 1.7698 days, where phases 0,1,... are defined to be at 
mid--primary eclipse. The curves show the fit for an eccentric orbit, to the primary and secondary
radial velocities. Points with error bars (error bars are masked by the symbol size in some cases) show the 
radial velocity measurements for the components of the system (primary: circles, secondary: squares). 
}
\label{RV;fig2}
\end{figure*}

\section{Spectroscopic analysis}
Double-lined spectroscopic binaries reveal two peaks, displacing back and forth, in the cross-correlation 
function (CCF)  between variable and the radial velocity template spectrum as seen in Fig.~1. The location of 
the peaks allows to measure of the radial velocity of each component at the time of observation. The cross-correlation 
technique applied to digitized spectra is now one of the standard tools for the measurement of radial velocities 
in close binary systems.

The radial velocities of V821\,Cas were obtained by cross--correlating of \'{e}chelle orders of 
V821\,Cas spectra with the spectra of the bright radial velocity standard stars $\alpha$ Lyr (A0V), 59 Her (A3IV) and 
$\iota$ Psc (F7V) (Nordstr\"om et al., 2004). For this purpose the IRAF task \textsf{fxcor} was used. 

Fig. 1 shows examples of CCFs of V821\,Cas near the first and second quadrature. The two non-blended peaks correspond 
to each component of V821\,Cas. We applied the cross-correlation technique to five wavelength regions with well-defined 
absorption lines of the primary and secondary components. These regions include the following lines: Si\,{\sc iii} 
4568 \AA, Mg\,{\sc ii} 4481 \AA, He\,{\sc i} 5016 \AA, He\,{\sc i} 4917 \AA, He\,{\sc i} 5876 \AA. The stronger CCF 
peak corresponds to the more massive component that also has a larger contribution to the observed spectrum. To better 
evaluate the centroids of the peaks (i.e. the radial velocity difference between the target and the template), we 
adopted two separate Gaussian fits for the case of significant peak separation.

The radial velocity measurements, listed in Table 2 together with their standard errors, are weighted means of
the individual values deduced from each order (see, e.g., Frasca et al. 2006). The observational points and their 
error bars are displayed in Fig. 2 as a function of orbital phase as calculated by means of the ephemeris based on 
the photometric times of the primary eclipse described in De\v{g}irmenci et al. (2007).

The first detailed solution of both radial velocity curves of V821\,Cas components is presented in this
study. We found the semi-amplitude of the more massive, more luminous component to be K$_1$=120$\pm$2 
km s$^{-1}$ and K$_2$=150$\pm$2 km s$^{-1}$ for the secondary component.

\begin{figure*}
\center
\includegraphics[width=15cm]{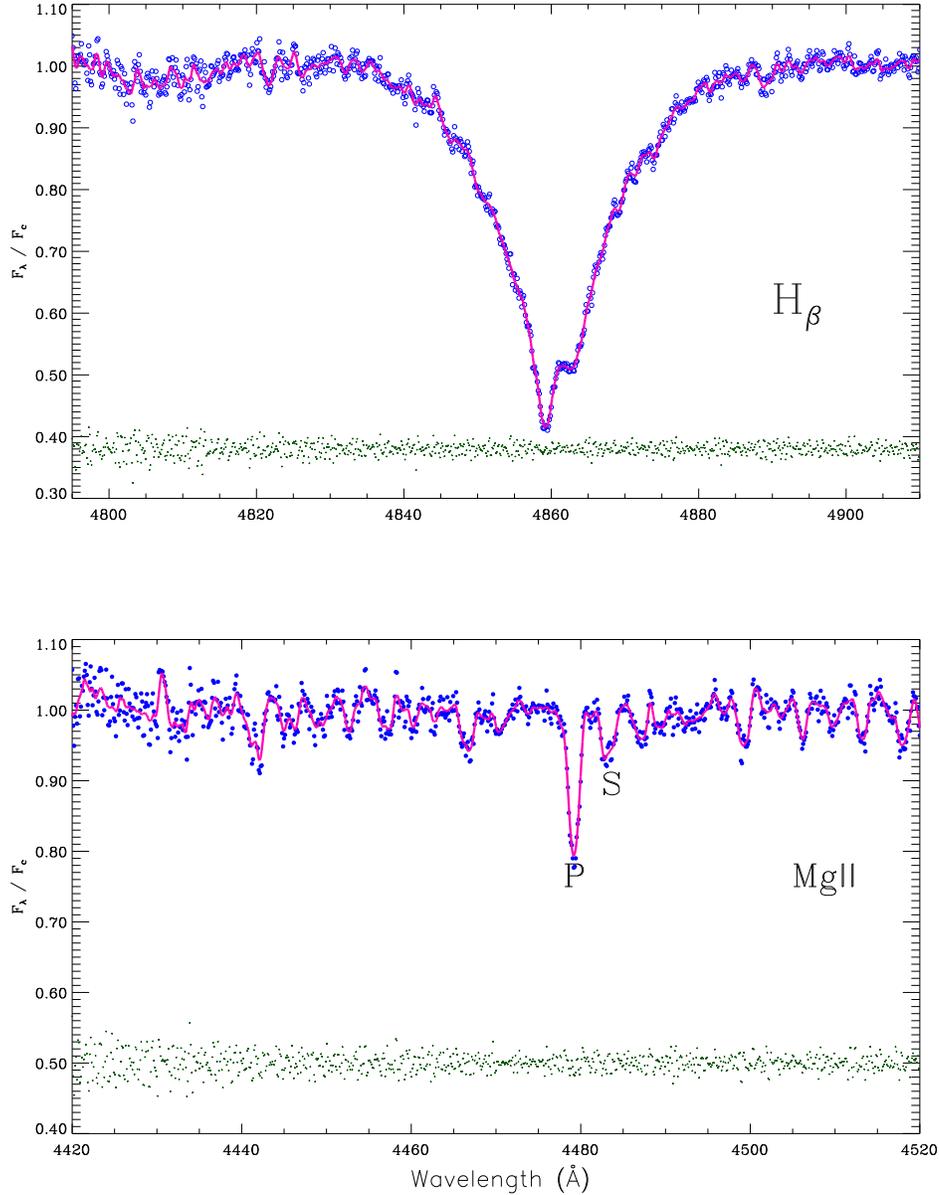}
\caption{Observed spectrum of V821\,Cas (large dots) in the $\lambda$4861 H$_{\beta}$ region (upper panel) and 
Mg\,{\sc ii} $\lambda$4481 (bottom). The synthetic spectrum (A1V+A4V) is displayed with continuous line in the same 
boxes. The differences (observed-synthetic, shifted) are plotted in the bottom of each panel.  
} 
\label{Vrot;fig3}
\end{figure*}

\subsection{Spectral classification}
We have used our spectra to classify the primary component of V821\,Cas. For this purpose we have degraded the spectral
resolution from 20\,000 to 3\,000, by convolving them with a Gaussian kernel of the appropriate width, and we have measured
the equivalent width ($EW$) of photospheric absorption lines useful for the spectral classification. 
We have followed the procedures of Hern\'andez et al. (2004), choosing hydrogen and helium lines in the blue-wavelength 
region, where the contribution of the secondary component to the observed spectrum is negligible. From several spectra we 
measured  $EW_{\rm H\gamma}=11.3\pm 0.9$\,\AA, $EW_{\rm H\alpha}=9.89\pm 0.11$\,\AA, 
$EW_{\rm H\beta}=10.79\pm 0.13$\,\AA, and $EW_{\rm MgII\lambda 4481}=0.31\pm 0.04$\,\AA.

From the calibration relations $EW$--Spectral-Type of Hern\'andez et al. (2004), we have derived a spectral
type A1V with an uncertainty of about 1.0 spectral subclass. The effective temperature deduced from the calibrations 
of Drilling \& Landolt (2000) or de Jager \& Nieuwenhuijzen (1987) is about 9\,400\,K (A1V). The spectral-type uncertainty 
leads to a temperature error of $\Delta T_{\rm eff} \approx 400$\,K.

\setcounter{table}{1}
\begin{table}
\centering
\caption{Radial velocities of the V821\,Cas' components. The columns give the heliocentric Julian date, the
orbital phase (according to the ephemeris given by De\v{g}irmenci et al. 2006), the radial velocities 
of the two components with the corresponding errors, and the average S/N of the spectrum. }
\begin{tabular}{@{}ccrcrcc@{}}
\hline
\textsf {HJD} & Phase& \multicolumn{2}{c}{Star 1 }& \multicolumn{2}{c}{Star 2 } &  $<S/N>$ \\
  2\,453\,000+ &  & \textsf{{\bf V$_p$}} & $\sigma$ & \textsf{{\bf V$_s$}} & $\sigma$&  \\
\hline
53953.6017  &0.3363	&-93.1  &6.1     & 124.8  &  9.3  & 88$^a$	\\
53955.6040  &0.4677	&-3.3   &3.2     & ---    &  ---  & 76		\\
53957.5655  &0.5761	&75.0   &6.4     & -88.0  &  8.8  & 80		\\
53958.5906  &0.1553	&-111.6 &8.3     & 148.1  &  11.6 & 97$^a$  \\
53970.5469  &0.9112	&61.0   &6.3     & -82.4  &  12.1 & 75	 	\\
53973.5377  &0.6012	&84.0   &4.8     & -104.0 &  8.1  & 95$^a$	\\
53975.5607  &0.7443	&107.5  &4.8     & -135.9 &  8.5  & 106$^a$	\\
53980.5290  &0.5516	&50.6   &4.9     & -73.1  &  9.9  & 90		\\
53981.5213  &0.1123	&-75.3  &3.6     & 100.8  &  7.1  & 90$^a$	\\
53982.5825  &0.7119	&108.0  &2.7     & -127.0 &  7.3  & 110$^a$	\\
53983.5405  &0.2533	&-133.0 &2.9     & 159.4  &  6.6  & 113$^a$	\\
53983.6089  &0.2919	&-121.5 &2.6     & 152.4  &  3.7  & 95		\\
53984.5390  &0.8175	&100.4  &3.0     & -123.2 &  5.2  & 97		\\
53984.6180  &0.8621	&89.3   &1.8     & -104.9 &  7.1  & 101$^a$	\\
53985.5492  &0.3883	&-53.4  &3.4     & 76.1   &  6.0  & 100$^a$	\\
53985.6109  &0.4231	&-40.2  &3.9     & 42.8   &  5.6  & 72		\\
\hline
\end{tabular}
\begin{list}{}{}
\item[$^a$]{\small Used also for rotational velocities ($v\sin i$) measurements.}
\end{list}
\end{table}

\subsection{Reddening}
The measurement of reddening is a key step in determining the absolute temperature scale (and therefore the distance) of 
eclipsing binaries. In addition to moderate distance determined by the Hipparcos mission, some reddening is expected for 
V821\,Cas due to its low galactic latitude ($l$=115$^{\degr}.10$, $b=-8^{\degr}.40$).

Our spectra cover the interstellar Na{\sc i} (5890 and 5896 \AA) doublets, which is excellent estimators of the reddening as 
demonstrated by Munari \& Zwitter (1997). They calibrated a tight relation linking the Na {\sc i} D2 (5890 \AA) and K{\sc i} 
(7699 \AA) equivalent widths with the E(B-V) reddening. On spectra obtained at quadratures, lines from both components 
are un-blended with the interstellar ones, which can therefore be accurately measured. We derive an equivalent width of 
0.34$\pm$0.02 \AA~for only Na{\sc i}, which corresponds to E(B-V)= 0.147$\pm$0.011 mag. K{\sc i} interstellar line is 
out of our spectral range as given in wavelength region in previous section.

\subsection{Rotational velocity}
The width of the cross-correlation profile is a good tool for the measurement of $v \sin i$ (see, e.g., 
Queloz et al. 1998). The rotational velocities ($v \sin i$) of the two components were obtained by 
measuring the FWHM of the CCF peaks in nine high-S/N spectra of V821\,Cas acquired close to the 
quadratures, where the spectral lines have the largest Doppler-shifts. In order to construct a 
calibration curve FWHM--$v \sin i$, we have used an average spectrum of HD~27962, acquired with 
the same instrumentation. Since the rotational velocity of HD~27962 is very low but not zero 
($v \sin i$ $\simeq$11 km s$^{-1}$, e.g., Royer, Zorec \& Fremat 2004 and references therein), it could be 
considered as a useful template for A-type stars rotating faster than $v \sin i$ $\simeq$ 10 
km s$^{-1}$. The spectrum of HD~27962 was synthetically broadened by convolution with rotational 
profiles of increasing $v \sin i$ in steps of 5 km s$^{-1}$ and the cross-correlation with the original 
one was performed at each step. The FWHM of the CCF peak was measured and the FWHM-$v \sin i$ 
calibration was established. The $v \sin i$ values of the two components of V821\,Cas were derived 
from the FWHM of their CCF peak and the aforementioned calibration relations, for a few wavelength 
regions and for the best spectra. This gave values of 70$\pm$1 km s$^{-1}$ for the primary star 
and 57$\pm$1 km s$^{-1}$ for the secondary star. 

We classified the spectral type of the primary component as an A1 type main-sequence star. The secondary component appears
to be an A4V star (see \S~5). The spectral types of the standard stars  $\alpha$ Lyr and 59\,Her are very close
to the primary and secondary component of the V821\,Cas, respectively. For the construction of the synthetic spectrum of the
system the spectra of the standard stars, obtained with the same  instrumentation, have been rotationally broadened by
convolution with the appropriate rotational profile and then co-added, properly weighted by using physical parameters
($T_1$, $T_2$, $R_1$, $R_2$, $vsini_{12}$) of the components as input parameters and, then, Doppler-shifted according 
to the radial velocities of the components. In Fig. 3 we show the simulation of the spectrum of V821\,Cas with the standard stars.

\section{Ephemeris and apsidal motion}
V821\, Cas has a moderate eccentric orbit. All available times of minimum light covering about 3100 orbital cycles were collected 
by De\v{g}irmenci et al. (2007). Their analysis of the O-C residuals yields for the first time a period of apsidal motion to be 158 yr.

\begin{figure*}
\includegraphics[width=16cm]{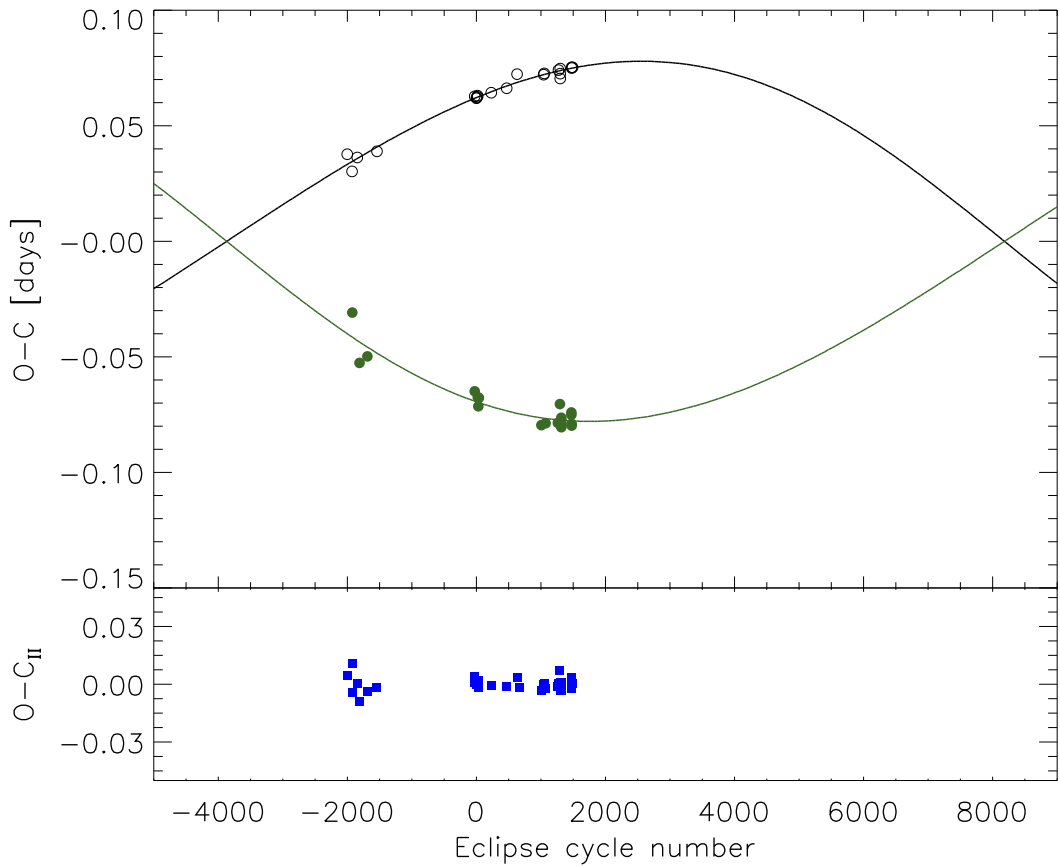}
\caption{Representation of the best-fitting apsidal motion parameters. The upper panel shows the observed
times of primary (open circles) and secondary (filled circles) minima, continuous line are best-fitting 
curves of primary and secondary minima. The lower panel shows the residual of the O-C values from the fitted curves.}
\label{o-c;fig4}
\end{figure*}

All available times of minimum light were collected from literature are listed in Table 4 together with the new 
timings obtained in this study. The O-C residuals indicate the differences between the observed and calculated times 
obtained with linear ephemeris,

\begin{equation}
Min I(HJD)=2\,451\,767.3504+1^d.7697397 \times E
\end{equation}

The method of Gimenez \& Garcia-Pelayo (1983) was adopted to carry out the apsidal motion 
parameters with the use of differential-correction approximations. The method is successively relates the derived apsidal 
motion parameters to the observed times of minima. We used some subroutines in the analysis of the O-C values, supplied 
by \"O. L. De\v{g}irmenci. Two secondary and one primary eclipse timings given by Otero (2005) were omitted from the analysis due to the large 
deviations from the others. These timings were estimated from very few observations, either in ascending or descending 
branch of the eclipses. Assuming the orbital inclination $i\sim83^{\circ}$ from the light curve analysis (see next section) 
and using the residuals in Table 4 we obtained the parameters of apsidal motion with a weighted least-square solution. 

\begin{table*}
\caption{Times of minimum light of V821\,Cas. The O-C values refer to the difference between the 
observed and calculated values.}
\label{O-C values}
\begin{tabular}{cccrcccr}
\hline
   Cycle number & Minimum time   & O-C & Ref. &Cycle number & Minimum time   & O-C & Ref. \\
                & (HJD-240\,0000)&     &      &              & (HJD-240\,0000)&     &\\
  \hline
-2002.0  & 48224.3680 &   0.0377   &1	&672.5    & 52957.4215 &  -0.0762   &4 \\
-1927.0  & 48357.0910 &   0.0303   &1	&1051.0   & 53627.4166 &   0.0727   &4 \\
-1923.5  & 48363.2240 &  -0.0308   &1	&1072.5   & 53665.3146 &  -0.0787   &4 \\
-1846.0  & 48500.4459 &   0.0363   &2	&1006.5   & 53548.5110 &  -0.0796   &7 \\
-1811.5  & 48561.4130 &  -0.0526   &1	&1042.0   & 53611.4884 &   0.0721   &7 \\
-1689.5  & 48777.3240 &  -0.0498   &1	&1262.5   & 54001.5652 &  -0.0786   &7 \\
-1541.0  & 49040.2190 &   0.0389   &1	&1272.0   & 54018.5304 &   0.0741   &7 \\
-197.5   & 51417.6940**& -0.1306   &1	&1294.5   & 54058.2050 &  -0.0704   &8 \\
-179.0   & 51450.6030**&  0.0382   &1	&1299.0   & 54066.3097 &   0.0704   &8 \\
-162.5   & 51479.6370**& -0.1285   &1	&1299.0   & 54066.3118 &   0.0725   &8 \\
-26.5    & 51720.3851 &  -0.0649   &3	&1299.0   & 54066.3140 &   0.0747   &8 \\
-26.0    & 51721.3976 &   0.0627   &3	&1315.5   & 54095.3595 &  -0.0804   &8 \\
0.0      & 51767.4100 &   0.0619   &4	&1315.5   & 54095.3601 &  -0.0798   &8 \\
4.0      & 51774.4893 &   0.0622   &4	&1315.5   & 54095.3610 &  -0.0789   &8 \\
17.0     & 51797.4962 &   0.0625   &4	&1315.5   & 54095.3625 &  -0.0774   &8 \\
17.0     & 51797.4967 &   0.0630   &4	&1315.5   & 54095.3636 &  -0.0763   &8 \\
21.5     & 51805.3300 &  -0.0675   &4	&1471.5   & 54371.4442 &  -0.0751   &8 \\
29.5     & 51819.4840 &  -0.0714   &4	&1471.5   & 54371.4452 &  -0.0741   &8 \\
38.5     & 51835.4153 &  -0.0678   &4	&1475.5   & 54378.5192 &  -0.0790   &8 \\
230.0    & 52174.4524 &   0.0643   &4	&1476.0   & 54379.5584 &   0.0753   &8 \\
469.0    & 52597.4220 &   0.0662   &5	&1476.5   & 54380.2882 &  -0.0798   &8 \\
630.0    & 52882.3561 &   0.0723   &6	&1477.0   & 54381.3278 &   0.0750   &8 \\
672.5    & 52957.4215 &  -0.0762   &4  	&1490.0   & 54404.3347 &   0.0753   &8 \\ 
\hline
\end{tabular}
\begin{list}{}{}
\item[$^{**}$]{\small Rejected from the fit owing to a large O-C value.} 
\item[Ref:]{\small (1) Otero (2005), (2) ESA (1997),(3) Bulut \& Demircan (2003), 
(4) De\v{g}irmenci et al. (2003, 2007), (5) Ak \& Filiz (2003), (6) Bak{\i}\c{s} 
et al. (2003), (7) Brat et al. (2007), (8) This study.}
\end{list}
\end{table*}

\begin{table}
  \caption{Apsidal motion parameters for V821\,Cas.}
  \label{apsidal motion parameters}
  \begin{tabular}{lr}
  \hline
Anomalistic Period $P_a$						&1.769813$\pm$0.000040		\\
Reference minimum time $T_0$ HJD				&51\,767.3481$\pm$0.0008	\\
Periastron longitude at $T_0$,$\omega^{\circ}$	&148$\pm$4					\\
Apsidal motion rate $\dot{\omega}$ cycle$^{-1}$	&0.0149$\pm$0.0023   		\\
Orbital eccentricity $e$						&0.138$\pm$0.011			\\
Apsidal motion rate $U$ yr						&117$\pm$19					\\
  \hline
  \end{tabular}
\end{table}

The apsidal motion resulting from the fit is $\dot{\omega}$=0$^{\circ}$.0149 $\pm$0$^{\circ}$.0023 cycle$^{-1}$, which 
is significant at the 6.5$\sigma$ level. The sidereal and anomalistic periods ($P_s$ and $P_a$), as well as the apsidal motion 
period $U=118\pm19$ yr are listed in Table 4. A plot of the O-C deviations of the times of minimum from the linear 
terms of the apsidal motion is shown in Fig. 4 along with the predicted deviations. In the bottom panel of Fig. 4 we show the
differences between observed and computed timings, taking into account apsidal motion. The apsidal motion parameters given in 
Table 4 are only preliminary results because the data we used cover only 14 \% of the apsidal motion period.

\begin{figure*}
\includegraphics[width=16cm]{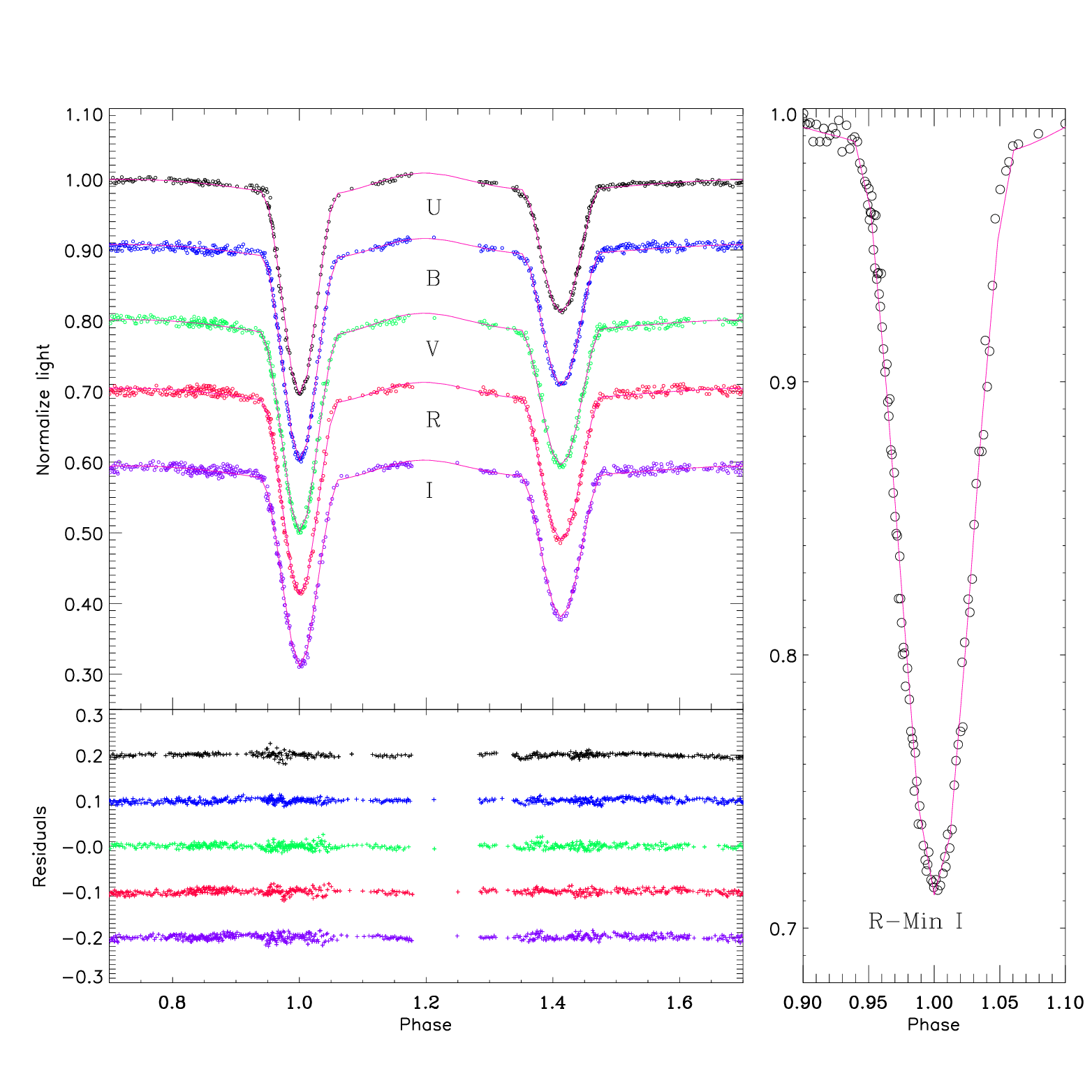}
\caption{Observed phased light curves of V821\,Cas with the best-fitting Phoebe model light curves. For clarity the 
{\em UBVRI} light curves and residuals of the fit (lower panel) have been offset by constant amounts as shown in figure. The right panel expands the region 
near the primary minimum for the R-band.}
\label{LC;fig5}
\end{figure*}

\section{Simultaneous light curves and radial velocities analyses}
In order to obtain the values of the main physical parameters of the binary components and to determine  
the orbital inclination ($i$) of the system (necessary to deduce stellar masses from the $M_{1,2}\sin^3 i$ 
derived from the RV curves) we applied a numerical eclipsing binary model to the {\em UBVRI} observations. We 
choose the Wilson--Devinney (W--D) code implemented into the PHOEBE package tool by Prsa \& Zwitter (2005) 
for the LC and differential correction (DC) fits.

The code was set in Mode--2 for detached binaries with no constraints on the potentials. The effective temperature 
($T_{1}$) of the primary star was fixed at 9400$\pm$400 K defined in \S 3. The simplest considerations were
applied for the parameters of the stars in the model, that is, stars were considered as black bodies, approximate
reflection model (MREF=1) was adopted, and no third light or spots were included. Gravity--darkening exponents
$g_1=g_2$ = 1 and bolometric albedos $Alb_{1}=Alb_{2}=1$ were set for radiative envelopes. We used square root
limb--darkening law. Bolometric limb-darkening coefficients were taken from Van Hamme (1993).

The mass ratio, $q$=$M_2$/$M_1$, is very important parameter in the light curve analysis, because the WD code is based on 
Roche geometry which is sensitive to this quantity. The mass ratio of 0.795 determined from the radial velocities 
was kept as a fix value. The iterations were carried 
out automatically until convergence and a solution was defined as the set of parameters for which the differential corrections 
were smaller than the probable errors. The light curves were analyzed individually and the weighted means of the parameters $i$, 
T$_2$, $\Omega_1$, $\Omega_2$, $r_1$ and $r_2$ were computed. Our final results are listed in Table 5 and the computed light curves are shown as 
continuous lines in Fig. 5. The uncertainties assigned to the adjusted parameters are the internal errors provided 
directly by the Wilson-Devinney code. In the bottom panel of Fig.5 the residuals between observed and computed fluxes are also 
plotted. The residuals reveal that the binary model may represent the observed light curves successfully.

\begin{table}
  \caption{Results from the simultaneous solution of {\em UBVRI} band light curves of V821\,Cas.}
  \label{parameters}
  \begin{tabular}{lr}
  \hline
   Parameter & Value\\
   \hline
   $i (^{\circ})$ 				& 82.6$\pm$0.1  			\\
   $e (^{\circ})$ 				& 0.127$\pm$0.007  	\\
   $\omega (^{\circ})$ 			& 155$\pm$4  			\\
   $T_{1}$ (K) 					& 9400[\textsf{Fix}] 		\\
   $T_{2}$ (K)					& 8600$\pm$17 		\\
   $\Omega_{1}$ 				& 5.067$\pm$0.014 	\\
   $\Omega_{2}$ 				& 6.693$\pm$0.035 	\\
   $q_{spec}$ 					& 0.795					\\
   $L_{1}/(L_{1+2})_U$ 			& 0.825$\pm$0.001  	\\
   $L_{1}/(L_{1+2})_B$ 			& 0.794$\pm$0.003  	\\
   $L_{1}/(L_{1+2})_V$ 			& 0.801$\pm$0.004 	\\
   $L_{1}/(L_{1+2})_{R}$ 		& 0.790$\pm$0.004  	\\
   $L_{1}/(L_{1+2})_I$ 			& 0.781$\pm$0.008  	\\
   $r_1$						& 0.2434$\pm$0.0013  	\\
   $r_2$						& 0.1466$\pm$0.0017  	\\
   $\Delta \phi$				&-0.0424$\pm$0.0005 		\\
   $\chi^2$						& 0.002  					\\
  \hline
  \end{tabular}
\end{table}

\begin{table}
 \setlength{\tabcolsep}{2.5pt} 
  \caption{Fundamental parameters of V821\,Cas.}
  \label{parameters}
  \begin{tabular}{lcc}
  \hline
  & \multicolumn{2}{c}{V821\,Cas} 		\\
   Parameter 						& Primary	&	Secondary				\\
   \hline
   Spectral Type					& A1V$\pm$1  	& A4V$\pm$1    			\\
   $a$ (R$_{\odot}$)				&\multicolumn{2}{c}{9.496$\pm$0.099}	\\
   $V_{\gamma}$ (km s$^{-1}$)		&\multicolumn{2}{c}{-0.05$\pm$0.01} 	\\
   $i$ ($^{\circ}$)					&\multicolumn{2}{c}{82.6$\pm$0.1} 		\\
   $q$								&\multicolumn{2}{c}{0.795$\pm$0.017}	\\
   Mass (M$_{\odot}$) 				& 2.046$\pm$0.067 & 1.626$\pm$0.058		\\
   Radius (R$_{\odot}$) 			& 2.311$\pm$0.028 & 1.392$\pm$0.022		\\
   $\log~g$ ($cgs$) 				& 4.021$\pm$0.007 & 4.362$\pm$0.012		\\
   $T_{eff}$ (K)					& 9400$\pm$400	& 8600$\pm$400      	\\
   $(vsin~i)_{obs}$ (km s$^{-1}$)	& 70$\pm$1		& 57$\pm$1       		\\
   $(vsin~i)_{calc.}$ (km s$^{-1}$)	& 66$\pm$1		& 40$\pm$1		       	\\
   $\log~(L/L_{\odot})$				& 1.58$\pm$0.06	& 0.98$\pm$0.01       	\\
   $d$ (pc)							& \multicolumn{2}{c}{260$\pm$12}			\\
   $J$, $H$, $K_s$ (mag)$^{*}$		& \multicolumn{2}{c}{7.978$\pm$0.018, 8.007$\pm$0.027, 7.956$\pm$0.027}	\\
$\mu_\alpha cos\delta$, $\mu_\delta$(mas yr$^{-1}$)$^{**}$ & \multicolumn{2}{c}{-2.51$\pm$0.59, -5.95$\pm$0.62} \\
$U, V, W$ (km s$^{-1}$)  & \multicolumn{2}{c}{4.77$\pm$0.70, 1.07$\pm$0.34, -6.82$\pm$0.80}\\ 
\hline  
  \end{tabular}

\medskip
{\rm *{\em 2MASS} All-Sky Point Source Catalogue (Cutri et al. 2003)} \\ 
{\rm **Newly Reduced Hipparcos Catalogue (van Leeuwen 2007)} \\ 
\end{table}

\section{Discussion and conclusions }
\subsection{Absolute dimensions and distance to the system}
Combination of the parameters obtained from light curves and RVs yield the absolute dimensions of the system, which are 
presented in Table 6. The standard deviations of the parameters have been determined by JKTABSDIM\footnote{This can 
be obtained from http://www.astro.keele.ac.uk/$\sim$jkt/codes.html} code, which calculates distance and other physical 
parameters using several different sources for bolometric corrections (Southworth et al. 2005). The radii of the 
components were estimated with uncertainties of 1.2 \% and 1.6\%. However, the uncertainties on the masses, being 
3.3 \% and 3.6 \%, are slightly larger than the criteria of Andersen (1991) for the necessary precision of absolute 
dimensions of stars to be used for comparison with theoretical models. 

An inspection of the temperatures, masses and radii of the component stars reveals a binary system composed of two main-sequence 
stars. The temperature $T_{eff1}=9400$ K, mass $M_{1}=2.04$ M$_{\odot}$ and radius $R_{1}=2.31$ R$_{\odot}$ of the primary are 
consistent with the spectral type of A1V, and the temperature $T_{eff1}=8450$ K, mass $M_{2}=1.62$ M$_{\odot}$ and radius 
$R_{2}=1.39$ R$_{\odot}$ of the secondary are consistent with an A4V spectral type star.

The colour excess $E(B-V)$ for a star may also be determined from the wide-band photometry. For calculation the reddening-free 
$Q-parameter$ we used the well-known equation below,

\begin{equation}
Q = (U-B) - [E(U-B)/E(B-V)] \times (B-V),
\end{equation}
for stars from O9 to A2. The average ratio of colour excesses for stars from O8 to A2 was adopted 
as $[E(U-B)/E(B-V)]=0.72\pm0.03$ (Hovhannessian 2004). The standard relation between $Q_{UBV}$ and $(B-V)_{0}$ was also given 
in Hovhannessian's study. For calculation of the total absorption in the visual magnitude the following relation, the ratio of 
selective-to-total extinction in the $V$ band,

\begin{equation}
[A_{V} / E(B-V)] = 3.30 + 0.28(B-V)_{0} + 0.04 E(B-V),
\end{equation}
given by Drilling \& Landolt (2001) was adopted.

Using the observed visual magnitude and color indexes given by Oja (1985) as V=8.26, U-B=0.07 and B-V=0.11 mag we 
compute the reddening-free parameter as $Q=-0.009$. Using the tables given by Hovhannessian (2004) we find the intrinsic colour 
index of $(B-V)_{0}=-0.021$ mag and then the colour excess of $E(B-V)=0.131\pm0.020$ mag for the system. The colours of the system 
were the average of three measurements. We should mention that there is no clue about the date of observations obtained, i.e. orbital 
phase. This value is in agreement with the E(B-V)=0.147 determined from the spectra given in \S 3.2.

Using the two $E(B-V)$ values derived from photometric and spectroscopic data we calculated the de--reddening distance modulus
of the system. To estimate the bolometric magnitudes of the components we adopted M$_{bol}$=4.74 mag for the Sun. Using the 
bolometric corrections given by Drilling \& Landolt (2000) and Girardi et al. (2002) we estimate the distance to the system as 260 
and 274 pc, respectively, with an uncertainty of 12 pc. However, the average distance to the system is estimated to be 
206$^{+42}_{-29}$ pc from the trigonometric parallax measured by the Hipparcos mission.

\subsection{Internal structure}
Our detection of the apsidal motion rate of the V821\,Cas in \S 4 provides the opportunity to test the models of stellar internal 
structure. In addition to the classical Newtonian contribution, the observed rate of rotation of apsides includes 
also the contributions arising from General Relativity. The theory of General Relativity 
estimates the relativistic contribution to observed rate as the following Einstein formula:

\begin{equation}
\dot{\omega_{rel}}=5.45\times10^{-4}\frac{1}{1-e^2} (\frac{M_1+M_2}{P_a})^{2/3}
\end{equation}
where M$_i$($i=1,2$) denotes the individual masses of the components in solar mass, $e$ is the orbital eccentricity (from radial
velocity analysis) and $P_a$ is the anomalistic period of the system in days. Use of equation (4) for V821\,Cas yields the relativistic 
contribution of 0$^{\degr}$.0009 cycle$^{-1}$ which is only 6 per cent of the observed rate. Removing the relativistic contribution
from the observed rate of apsidal motion, one can calculate the observational average value of internal structure constant
($k_{2obs}$) using the following formula:

\begin{equation}
\bar{k}_{2obs}=\frac{1}{c_{21}+c_{22}} \frac{\dot{\omega}}{360},
\end{equation}
where c$_{2i}$($i=1,2$) are the functions of the orbital eccentricity, fractional radii and masses of the components and ratio 
of the rotational velocities of the component stars to the Keplerian velocity. The observed average value of $\log k_2$ is found
to be -2.56$\pm$0.07 from Eq. (5). The theoretical internal structure constants for the components ($\log~k_{2theo1,2}$) are taken from the 
theoretical  calculations of Claret (2004). Among the tabulated values, interpolation for the masses and $\log~g$ of the 
component star V821\,Cas yielded -2.48 and -2.37 for the primary and secondary components, respectively. The mean theoretical value of 
internal structure constant was computed as $\log~k_{2theo}$=-2.43 which is in agreement with that of observed within 2$\sigma$ 
level. The internal structure constant obtained from the O-C analysis is smaller by 35 \% than predicted from the theory. The present 
analysis shows that the component stars are more concentrated in mass than predicted by theoretical calculations.

\begin{figure*}
\includegraphics[width=12cm]{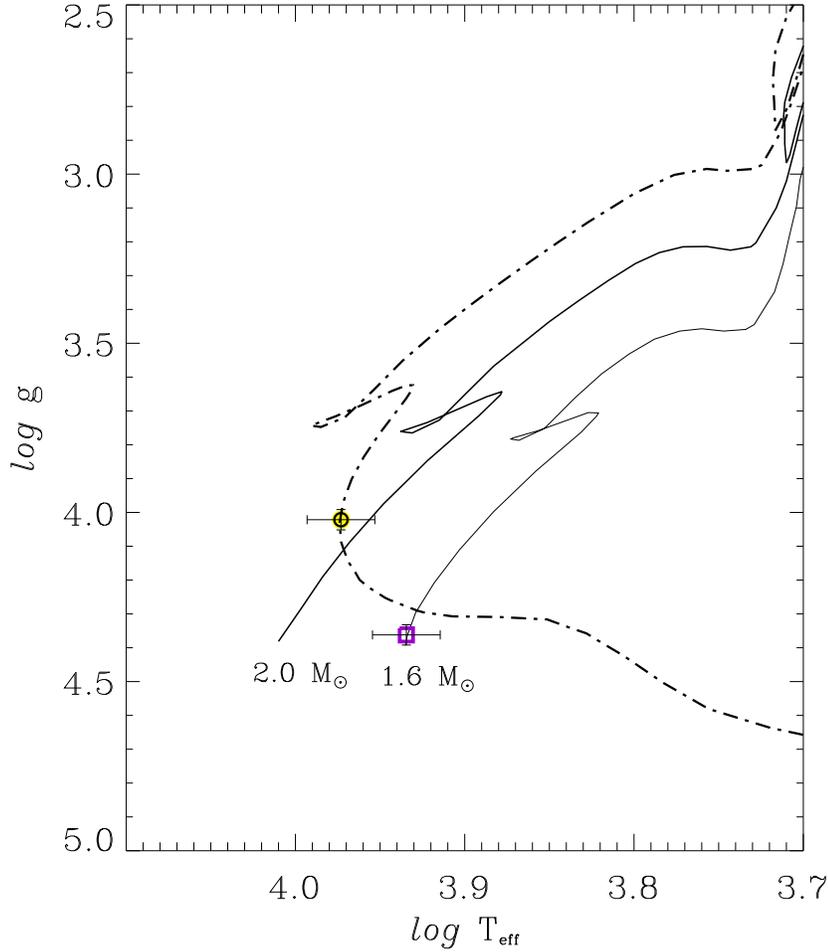}
\caption{Location of the two stellar components of V821\,Cas in $\log~T_{eff}$-$~\log~g$ diagram, together with evolutionary 
models for the masses of 1.6 and 2.0 M$_{\odot}$ (Girardi et al. 2002) and the isochrones of an age $t$=5.6$\times$10$^8$ 
 years (long-dashed-dotted line). The open circle corresponds to the primary and the open square to the secondary.
}
\label{evrim;fig6}
\end{figure*}

\subsection{Evolutionary stage and age of the system}
In Fig. 6, we plot the location of V821\,Cas stellar components in $\log~T_{eff}-\log~g$ diagram. The evolutionary tracks for masses 
1.6, 2.0 M$_{\odot}$ and $z=0.008$ (Girardi et al. 2002) are also shown in this figure. Higher metallicity tracks given by Girardi 
et al. do not match with the observed properties of the components. 
An isochrone corresponding to an age of 5.6$\times10^{8}$ yr, computed from Girardi et al. (2002)\footnote{We computed the isochrones using 
the facility available in the web site: http://stev.oapd.inaf.it/$\sim$lgirardi/cgi-bin/cmd} is also shown in Fig. 6. The effective 
temperature of the primary component appears to be slightly higher than the model for 2 M$_{\odot}$. However, when the uncertainties of the 
effective temperatures of the components are taken into account it is seen that the deviations from the models are less than 1$\sigma$ 
level. The diagram suggests that both components of V821\,Cas are main sequence stars, relatively young with an age of about 560 Myr.

\subsection{Population type and kinematical analysis of V821\,Cas}
To study the kinematical properties of V821\,Cas, we used the system's center-of-mass' velocity, distance and proper motion values, which are given in 
Table 6. The proper motion data were taken from newly reduced Hipparcos catalogue (van Leeuwen 2007), whereas the center-of-mass velocity and 
distance are obtained in this study. The system's space velocity was calculated using Johnson \& Soderblom's (1987) algorithm. The U, V 
and W space velocity components and their errors were obtained and given in Table 6. To obtain the space velocity precisely the first-order galactic 
differential rotation correction was taken into account (Mihalas \& Binney 1981), and -1.08 and 0.65 kms$^{-1}$ differential corrections were applied 
to U and V space velocity components, respectively. The W velocity is not affected in this first-order approximation. As for the LSR correction, 
Mihalas \& Binney's (1981) values (9, 12, 7)$_{\odot}$ kms$^{-1}$ were used and the final space velocity of V821\,Cas was obtained 
as $S=19.36$ kms$^{-1}$. This value is in agreement with other young stars space velocities.

To determine the population type of V821\,Cas the galactic orbit of the system was examined. Using Dinescu et al. (1999) N-body code, the 
system's apogalactic ($R_{max}$) and perigalactic ($R_{min}$) distances were obtained as 9.09 and 8.07 kpc, respectively. Also, the maximum 
possible vertical separation from the galactic plane is $|z_{max}|$=50 pc for the system. When determining the ellipticity the following formula 
was used:

\begin{equation}
e=\frac{R_{max}-R_{min}}{R_{max}+R_{min}}.
\end{equation}
The ellipticity was calculated as $e=0.059$. This value shows that V821\,Cas is orbiting the Galaxy in an almost circular orbit and that the system 
belongs to the young thin-disc population.

\section*{Acknowledgments}
We thank Prof.\ G.\ Strazzulla, director of the Catania Astrophysical Observatory, and Dr. G.\ Leto, responsible 
for the M. G. Fracastoro observing station for their warm hospitality and allowance of telescope time for the 
observations. In addition, \"{O}\c{C} is grateful to all the people working at the Catania Astrophysical Observatory 
for creating a stimulating and enjoyable atmosphere and, in particular, to the technical staff of the OAC, namely 
P.\ Bruno, G.\ Carbonaro, A.\ Distefano, M.\ Miraglia, A.\ Miccich\`e, and G.\ Occhipinti, for the valuable support 
in carrying out the observations. We also thank to T\"UB\.ITAK for a partial support in using T40 with project 
number TUG-T40.000.111. EB{\.I}LTEM Ege University Science Foundation Project No:08/B\.{I}L/0.27 and Turkish 
Scientific and Technical Research Council for supporting this work through grant Nr. 108T210. We also thank to 
K. B. Co\c skuno\u glu for his contributions. We are also grateful to the anonymous referee whose comments 
helped to improve this paper. This research has been also partially supported by INAF and Italian MIUR. 
This research has been made use of the ADS and CDS databases, operated at the CDS, Strasbourg, France.

\label{lastpage}
\end{document}